# A Framework to model real-time databases

**Nizar Idoudi [1], Nada Louati [2], Claude Duvallet [1], Rafik Bouaziz [2], Bruno Sadeg [1] and Faiez Gargouri [2]**

[1] LITIS Laboratory, Faculty of Sciences and Techniques, Le Havre University, France
[2] MIRACL Laboratory, Higher Institute of Computer Science and Multimedia, Sfax University, Tunisia
[nizar.idoudi, claude.duvallet, bruno.sadeg]@ univ-lehavre.fr
[nada.louati, raf.bouaziz, faiez.gargouri]@ fsegs.rnu.tn

**Abstract:** *Real-time databases deal with time-constrained data and time-constrained transactions. The design of this kind of databases requires the introduction of new concepts to support both data structures and the dynamic behaviour of the database. In this paper, we give an overview about different aspects of real-time databases and we clarify requirements of their modelling. Then, we present a framework for real-time database design and describe its fundamental operations. A case study demonstrates the validity of the structural model and illustrates SQL queries and Java code generated from the classes of the model.*

**Keywords:** *Real-time database, Real-time objects, Object-relational model, UML Profile, UML CASE Tools.*

## 1. Introduction

Real-time databases have to deal with time-constrained data and time-constrained transactions. They are now being used for several applications such as space project control, process control, financial market and air traffic control systems. In each of these time-critical applications, data about the target environment must be continuously collected from the real-world and processed in a timely manner to generate real-time responses. A real-time database has two distinguishing features: the notion of temporal consistent data, and the ability to place real-time constraints on transactions. Some of its data must not only be logically consistent, but also temporally consistent, i.e., must closely reflect the current state of the controlled environment. But data are collected at discreet moments. Hence, they often represent an approximation of reality. As time advances continually, a real-time data value becomes less and less accurate, until the moment where it does not reflect any more the state of the environment. At this time point, we say that this data value is no longer temporally consistent. Temporal consistency can be measured in two ways: *absolute consistency* and *relative consistency* [22]. A data item, as the speed of an aircraft in an air traffic control system, is considered absolutely consistent if and only if its age is within a specified time interval. It should be often updated. The age represents the time duration between its timestamp and the current time. For example, the speed age value must not exceed five seconds; it verifies its absolute consistency constraint as long as it is no more than five seconds old. Relative consistency concerns data derived from other ones. For example, the lane of an aircraft is derived from the location and the altitude data items. Its temporal consistency depends of those of location and altitude data. So, real-time data are subdivided into two types: *sensor data* and *derived data* [24]. Sensor data are the data issued from sensors. Derived data are the data computed using sensor data (e.g., lane data).

Transactions in a real-time database environment are subdivided into two classes: *update transactions* and *user transactions* [22]. Update transactions are used to update the values of real-time data in order to reflect the state of the real world. They are executed periodically to update sensor data, or sporadically to update derived data. User transactions, representing user requests, arrive aperiodically. They may read or write non real-time data, but only read real-time data.

Real-time databases are therefore specific. Their design need appropriate concepts and tools which are not available under systemic or object oriented methods. UML, the most used nowadays, can not, in its standard form, satisfy the requirements of such design. Indeed, it is a general language used to model object-oriented applications across a wide range of domains. But its extension mechanism, based on the concepts of profile and stereotype, allows it, if enhanced, to support new concepts and tools and to become suitable for a particular domain needing specific software engineering activities. Recently, an UML profile for Modeling and Analysis of Real-Time and Embedded systems (MARTE) has been standardized by the OMG [17] [3]. However, the design of real-time databases differs from the design of conventional real-time systems. The designers of real-time databases must consider both temporal aspects of data and timing constraints of transactions [26] [9] [10]. The design of this kind of databases is thus performance-and semantic-dependent. It must consider factors such as sensor data, derived data and Quality of Data (QoD) management, temporal semantics in transaction scheduling algorithms, concurrency control protocols, disk caching, and buffer management protocols to meet the timing constraints defined by the real-time applications. Which concepts and tools are suitable to define the design of real-time databases? How to define and to implement an UML profile supporting these requirements? How to implement a real-time database model under a technical environment? The present paper deals with these problems and aims to bring contributions in real-time databases design.

The paper is organized as follows. Section 2 briefly covers related works. Section 3 describes our real-time object-oriented data model. Section 4 presents a set of stereotypes that allows our UML-RTDB profile to express real-time database features in a structural model. Section 5 quotes different mapping rules of an UML-RTDB diagram to an object relationship schema. Finally, the last section draws some conclusions.

## 2. Related work

There is a need to define an UML profile supporting real-time database requirements. Several UML approaches were already proposed to take into account the real-time system requirements, such as *UML-RT* [25], *RT-UML* [5], *UML-SDL* [11] and *ACCORD/UML* [13]. The basic concepts of *RT-UML* were integrated in the UML standard through the UML profile for Schedulability, Performance, and Time (denoted SPT profile) [18], which is recently replaced by the UML profile for Modeling and Analysis of Real-Time and Embedded Systems (MARTE) [17]. However, UML constructs used by these approaches do not support real-time database requirements. A real-time database is a database in which both the data and the operations upon the data may have timing constraints [22]. In fact, real-time databases have all requirements of traditional databases, such as the management of accesses to structured, shared and permanent data, but they also require management of time-constrained data and time-constrained transactions [2].

To the best of our knowledge, there is only one UML-based proposal for real-time databases modeling [4]. In their work, the authors have defined an UML package for specifying RTSORAC[1] [29] object, called *RT-Object*. However, the RT-Object package is based on the Extension Mechanisms package of UML1.3 which is a past standard. Furthermore, imprecise computation encapsulated within the RTSORAC object is defined in the context of *Epsilon Serializability* (on transactions) [23], and does not support the notion of QoD introduced in [1]. The QoD concept allows a robust and controlled behavior of real-time databases during transient overloads, based on *Feedback Control Real-Time Scheduling* [16].

The framework proposed in this paper is distinguished by the fact that it supplies, like RT-Object, concepts and tools for real-time database modeling. But, unlike RT-Object, UML-RTDB, the profile of this framework, supports the *QoD* concept that we define for real-time attributes [9], on one hand, and it contains a set of stereotypes to express dynamic semantics of real-time attributes and real-time object features, on the other hand. These stereotypes are defined under UML.2.1.2 Profiles package [20]. In addition, UML-RTDB allows to specify two kinds of real-time attributes, *sensor* attributes and *derived* attributes, in order to satisfy the requirements of current real-time applications.

## 3. Real-time object model

The mostly used data model for real-time databases is the relational model [22]. However, due to the nature of many real-time applications, that must handle complex real-world objects with short deadlines, many researchers believe that the object-oriented model is more suitable and powerful than the relational model [12]. So, several research projects on real-time databases have adopted the object-oriented model for building their prototype systems [29] [27]. Our work is based on a particular object model, named *real-time object-oriented data model*, which incorporates time-constrained data and time-constrained transactions of real-time databases [9]. Thus, a real-time database is a collection of objects which are used to manage time-critical dynamic systems in the real world. Each object has some internal state which is protected by the object abstraction. The only way for a transaction to access an object is to invoque the methods defined by the class of this object.

In this section, we describe our data model and the characteristics of its components. We illustrate our proposal on an air traffic control system, which consists of a large collection of data describing the aircrafts, their flight plans, and environment data [15]. This includes flight information, such as aircraft identification, transponder code, altitude, position and speed, origin, destination, route and clearances. In our work, each aircraft in the airspace is modeled as a real-time object.

---

[1] RTSORAC: Real-Time Semantic Objects Relationships And Constraints.

## 3.1 Real-time object model

Real-time objects (RTO) are the real-time object-oriented database entities. They represent dynamic entities of time-critical dynamic applications in the real world [8]. We define a real-time object as an extension of the real-time object, as used in the ACCORD/UML approach [28] [6]. It encapsulates time-constrained data, time-constrained methods and concurrency control mechanisms. As shown in Figure 1, each real-time object is made of four components: (i) a set of real-time attributes, (ii) a set of real-time methods, (iii) a mailbox, and (iv) a local controller.

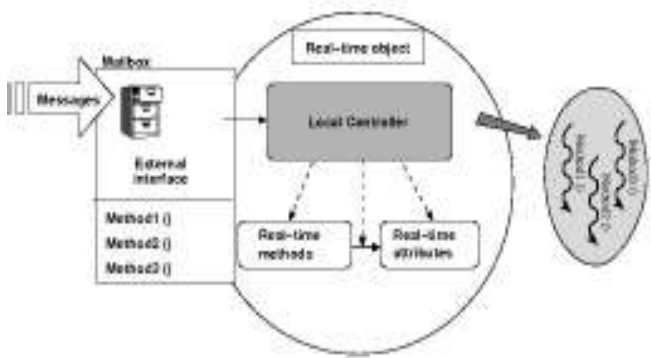

**Figure 1. Real-time object model.**

## 3.2 Mailbox

Because an object model tends to ensure modularity and encapsulation, it needs to use a unique method for exchanging information between two objects: *message passing*. In this context, all processing are triggered by message arrivals. The mailbox is used to store messages received by the RTO. A mailbox is attached to an object and is generally for a given object (instance). It is used only to store messages received by the object and waiting to be processed.

In most real-time applications, real-time constraints are attached to messages. So, communication between two objects is defined as a *Client-Server* relationship between these two objects, where message passing is a request of a service from the *Client* to *Server*. From this point of view, it is completely natural to adopt the usual rule of collaboration between a client and a supplier: "the client is always right", meaning that real time constraints are set by client and not by server. Thus, real-time constraints have to be attached to the service request, i.e., the message passing. In our model, we characterize each message by a *deadline* which must be met; otherwise the message will be rejected (*Firm*).

## 3.3 Local controller

Because of the dynamic nature of the real world, more than one transaction may send requests to the same real-time object. Concurrent execution of transactions allows several methods to run concurrently within the same object. To handle this essential property of real-time database systems, we associate to each real-time object a local concurrency control mechanism, named *local controller*, which manages the concurrent execution of its methods.

The local controller has to take into account each new request arrival. It has also to select the request to be executed according to the timing constraints of the different requests of the mailbox, in two cases: (1) when a thread becomes available; (2) when it is necessary to suspend the current request having the low priority to liberate a thread, because there is a request that cannot wait. We note that a request corresponds to a method of the target object.

Besides, the local controller must verify the concurrency constraints between the method of the selected request and the already running methods of the real-time object. If it detects a conflict, it aborts the request having the lowest priority. When a method terminates its execution, the corresponding thread is released and concurrency constraints are relaxed. But, if the service is periodic, the thread is not released, since it is allocated to support all periodic executions of the requested service. The local controller is thus in charge of mailbox management, scheduling constraints handling, concurrency constraints handling, and thread management.

## 3.4 Real-time attributes

Data objects are classified into either real-time or non real-time. A non real-time data is a classical data found in conventional databases, whereas a real-time data has a validity interval beyond which it becomes useless [22]. Real-time data change continuously to reflect the real world state (for example, the current temperature value). Each real-time data has a timestamp indicating the last observation of the real world state.

Our real-time data model is based on the model introduced in [22] and we associate to this model the notion of *maximum data error* (MDE) introduced in [1]. Thus, a real-time data is modeled by d = ($d_{value}$, $d_{timestamp}$, $d_{avi}$, $d_{mde}$), where $d_{value}$ represents the real world data value, $d_{timestamp}$ is the last time at which the attribute's value was updated, $d_{avi}$ is the absolute validity interval and $d_{mde}$ is the maximum amount of imprecision associated with the attribute's value.

To handle the property of our real-time data model, we define an attribute model, called *real-time attribute*, which incorporates fields to support logical constraints ($d_{value}$), temporal constraints ($d_{timestamp}$, $d_{avi}$) and *QoD* constraints ($d_{mde}$). As shown in Figure 2, each real-time attribute is characterized by *<N, CV, TS, VD, MDE>* [9] where:

| N (Name) | | | |
|---|---|---|---|
| CV (Current Value) | TS (TimeStamp) | VD (Validity Duration) | MDE (Maximum Data Error) |

**Figure 2. Structure of Real-time attribute.**

*Name (N):* is the name of the attribute.

*Current Value (CV):* is used to store the final attribute value captured by the related last update method. This field is used by the system to determine logical integrity constraints of the attribute value.

*TimeStamp (TS):* is used to store the last time at which the attribute's value was updated [22]. This characteristic is of the type *Time* which is supported by the UML2.0 standard. Access to the timestamp of an attribute is necessary for determining temporal consistency of this attribute. For example, in an aircraft object, there is an attribute for storing the altitude, called *Altitude*, to which a sensor regularly provides readings. This update is reported every twenty seconds. Thus the *Altitude* attribute is considered temporally inconsistent if the update does not occur within that time frame. There are many ways to define timestamps. In our model, the timestamp is the time when the value is produced. If the value is produced by a sensor device, then timestamp is the time when the value is read by the sensor. If the value is produced by a transaction, then timestamp is the time when the transaction completes. This field is used by the system to determine whether or not timing constraints have been violated.

*ValidityDuration (VD):* is used to store the length of *absolute validity interval* (denoted by *avi*) of the attribute value [22]. It represents the amount of time during which the attribute value is considered valid. This element is numeric and allows to determine, in association with *TS*, the *absolute consistency* of the attribute. A data item is considered absolutely consistent (fresh) with respect to time as long as the age of the data value is within a given interval [22]. For instance, the *Altitude* value is considered fresh if the current time is earlier than timestamp of Altitude followed by the length of the absolute validity interval (*avi*) of Altitude; i.e. {*Altitude.TS* ≤ *currenttime* < *Altitude.TS + Altitude.VD*}.

*Maximum Data Error (MDE):* is used to memorize the *absolute maximum data error* tolerated on the attribute value [1]. Currently, the demand for real-time database services has increased in most applications where it is desirable to execute transactions within their deadlines. They also have to use precise and fresh data in order to reflect the continuously changing external environment. However, in many applications, it seems to be difficult for transactions both to meet their deadlines and to keep the database consistent. To support these applications, the *QoD* concept is introduced in [1] to indicate that data stored in the database may have some deviation from its value in the real world. Thereby, data error, denoted *DE*, represents the deviation between the current data value and the updated value. The upper bound of the error is given by the *Maximum Data Error*. For instance, the maximum error on the *Speed* value is 5 km/h. This field allows the system to handle the unpredictable workload of the database by discarding sensor transactions where *DE ≤ Attribut.MDE*, and to enhance the freshness of data using Feedback Control Scheduling [1]. It has the same type as *CV* field.

We note that only the two first fields, *Name* and *Current Value*, are visible to the users. The other fields are used by real-time database system in order to maintain the temporal consistency of the real-time database.

### 3.5 Sensor and Derived attributes
Since real-time data is subdivided into two types: *sensor data* and *derived data* [24], we characterize the real-time object model by three types of attributes as follows:

| Aircraft | |
|---|---|
| Identifier<br>Destination | Classical |
| Direction<br>Location<br>Altitude<br>Speed | Sensor |
| Path<br>Lane | Derived |
| SetIdentifier ()<br>GetLane ()<br>GetSpeed () | Aperiodic |
| UpdateLocation ()<br>UpdateAltitude () | Periodic |
| ComputeLane ()<br>ComputePath () | Sporadic |

**Figure 3. Aircraft object.**

*Classical attributes:* they are used to store non real-time data [9]. As shown in Figure 3, we characterize the *Aircraft* object by two classical attributes, *Identifier* and *Destination*.

*Sensor attributes:* they are used to store sensor data which must be periodically updated in order to closely reflect the real world state of the application environment [9].

For example, we characterize the *Aircraft* object by four sensor attributes: *Direction*, *Location*, *Altitude*

and *Speed*. These attributes are periodically updated to reflect the state of an *Aircraft* instance.

***Derived attributes:*** they are used to store *derived data* that have to be calculated from sensor attributes [9]. We characterize the *Aircraft* object by two derived attributes: *Path*, which is calculated from *Direction* and *Location* values, and *Lane*, which is calculated from *Location* and *Altitude* values. We note that in this paper, we consider that a derived attribute is not characterized by an MDE field.

## 3.6 Real-time methods

We consider a method execution as a transaction which is composed of one or many sub-transactions (a method can call other methods) [14]. We classify the real-time object methods into three classes: *periodic methods*, *sporadic methods*, and *aperiodic methods*.

***Periodic methods:*** the temporal consistency of each sensor data is ensured by a sensor transaction, which periodically updates the value of the sensor data [24]. Thereby, we associate to each sensor attribute a *periodic* method, which periodically updates the values of the *Current Value* and the *TimeStamp* fields. We assume that a periodic method execution is a sensor transaction. This latter is defined as a write-only transaction which obtains the state of the environment and writes the sensed data to the database. We propose two types of timing constraints for periodic methods: *absolute timing constraints*, i.e. a deadline, and a *period*. The periodicity of the method execution is imposed by the validity time of each value of the sensor attribute. A periodic method must complete its execution before the deadline; otherwise the value to be written will be considered obsolete [22].

For example, we characterize the *Aircraft* object by a set of periodic methods such as:

*(1) UpdateAltitude():* which periodically carries out write operations of *Current Value* and *TimeStamp* fields of the *Altitude* attribute.

*(2) UpdateLocation():* which periodically performs write operations of *Current Value* and *TimeStamp* fields of the *Location* attribute.

***Sporadic methods:*** derived data are the data calculated from sensor data [24]. Thereby, we associate to each derived attribute a *sporadic* method, which sporadically calculates its value from sensor attributes. The access mode of the sporadic method to derive attribute value is always ``write''. Its timing constraints are also *deadline* and *periodicity*. The periodicity of the method execution depends on the considered update policy. In this paper, we consider a dynamic update policy as proposed in [7]. Then, we characterize the *Aircraft* object by the following sporadic methods:

*(1) ComputeLane():* which computes the *Lane* attribute value using *Location* and *Altitude* values.

*(2) ComputePath():* which computes the *Path* attribute value using *Location* and *Direction* values.

***Aperiodic methods:*** they include the remainder of methods that allow to read/write classical attributes and read only sensor and derived real-time attributes. User transactions typically arrive aperiodically. They do not write any temporal data, but they can read/write non temporal data and only read temporal data [24]. To include nested transactions in this object model, we assume that an aperiodic method execution is a user transaction which may invoke atomic operations or invoke other methods on other objects [14]. Operations represent the actions of the method. They include statements for conditional branching, looping, I/O, reads/writes of non real-time attributes, and reads/writes of real-time attributes, including their *current value*, *timestamp*, *validity duration* and *maximum data error* fields.

## 4. The UML-RTDB profile

In this section, we present an UML profile, entitled UML-RTDB, which is a specialized variant of the UML2.1.2 for real-time database applications. The main aim of our proposal is to supply, to the designers of real-time databases, UML extensions to support real-time database requirements. An UML extension is specified in the UML metamodel by a stereotype. This latter defines how an existing metaclass may be extended, and enables to use platform or domain specific terminology or notation, in addition to the ones used for the extended metaclass [20]. In our work, UML-RTDB stereotypes extend metamodel classes with specific sensor and derived attributes, specific periodic and sporadic operations and a specific real-time class that allow the design of class diagrams for real-time databases. We base our proposal on the *Extension relationship* proposed in UML2.1.2 *Profiles* package [20].

### 4.1 Real-time data type

As defined in section 3.4, each real-time attribute value is characterized by a timestamp, which indicates the time at which it was last updated. So, for each real-time attribute value corresponds a timestamp, which distinguishes it from other attribute's values. Thereby, as illustrated in Figure 4 for the attribute *Speed*, the values of the *VD* and *MDE* fields are the same for all real-time attributes. But, the values of the **TS** field change for each real-time attribute.

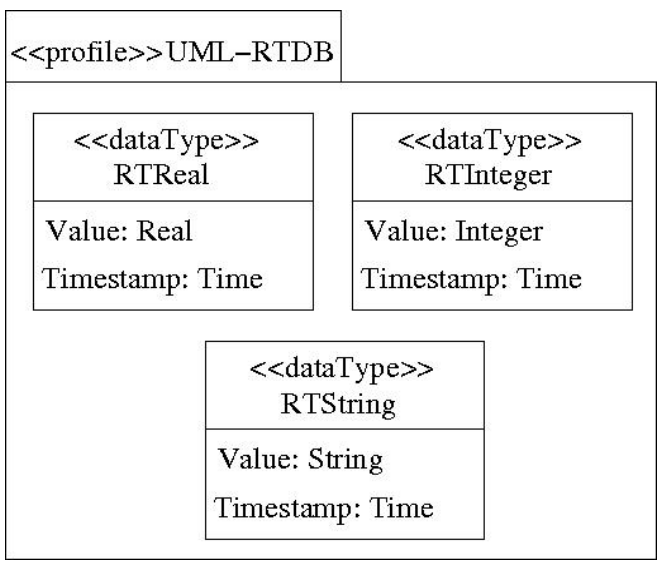

**Figure 4. Illustration of the dynamic semantic of a real-time attribute.**

For this reason, we define three new UML data types, called *RTInteger*, *RTReal* and *RTString*, that describe the type of real-time attribute values. As shown in Figure 5, each metaclasse is characterized by two properties:

(1) **Value:** it indicates the attribute value written by the related update method. It is of the type *Integer*, *Real* or *String* in the case of the metaclass *RTInteger*, *RTReal* or *RTString*, respectively.

(2) **Timestamp:** it indicates the last time at which the attribute's value was updated.

**Figure 5. Data types of UML-RTDB profile.**

In this work, we consider that time granularity is the *"Second"*. The designers of real-time databases can easily modify the values of stereotype properties according to the requirements of real-time applications.

### 4.2 Sensor and Derived stereotypes
Since a real-time attribute is either sensor or derived (cf. section 3.5), we define two stereotypes, <<*Sensor*>> and <<*Derived*>>, to declare respectively sensor attributes and derived attributes in the class diagrams. As shown in Figure 6, we define an abstract stereotype, called <<*RealTimeAttribute*>>, to factorize the Validity Duration property, which characterizes both <<*Sensor*>> and <<*Derived*>> stereotype. As for Maximum Data Error property, it characterizes only <<*Sensor*>> stereotype (cf. section 3.5).

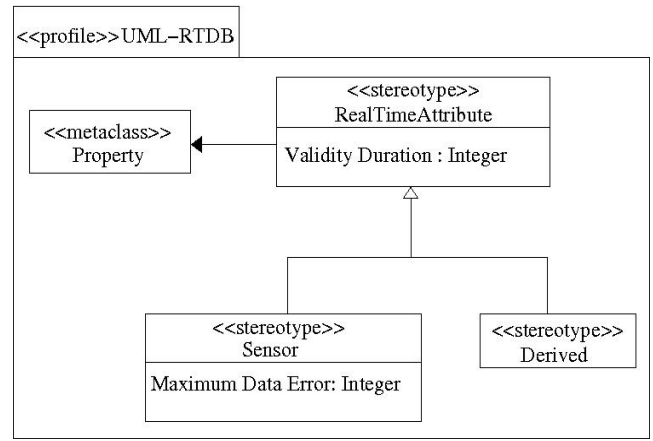

**Figure 6. Definition of Sensor and Derived stereotypes.**

### 4.3 Periodic and Sporadic stereotypes
We characterize the UML-RTDB profile by three stereotypes, <<*Periodic*>>, <<*Sporadic*>> and <<*Aperiodic*>>, to declare respectively periodic methods, sporadic methods and aperiodic methods in the class diagrams. As illustrated in Figure 7, we define an abstract stereotype, named <<*Update*>>, which generalizes these latter stereotypes. It is characterized by a Deadline, which indicates the last time by which the method execution must be completed. In addition, we characterize the <<*Periodic*>> stereotype by a Period in order to define the periodicity of the methods.

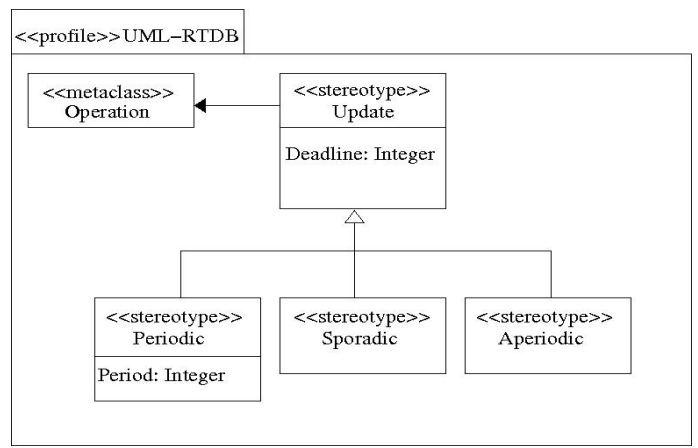

**Figure 7. Definition of Periodic and Sporadic stereotypes.**

### 4.4 Real-Time Class stereotype
A real-time database is by definition a database system. It then has queries, schemas, transactions, commit protocols, concurrency control support, and storage management [26]. So, the design of a real-time database has to take into account the management of all these components. That's why we define a <<*RealTimeClass*>> stereotype (cf. Figure 8) in order to deal with the time-constrained data, time-constrained operations, parallelism, and concurrency property inherent to real-time databases.

The <<*RealTimeClass*>> stereotype is added to classes in order to specify that their instances will encapsulate real-time data, real-time operations, and a local concurrency control mechanism.

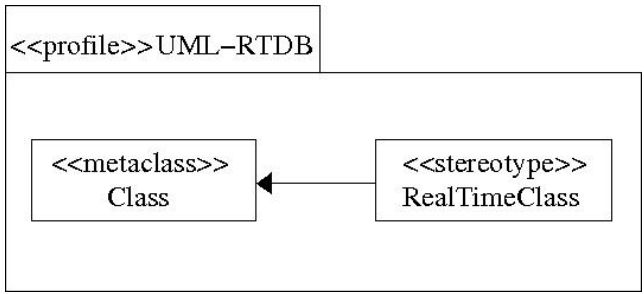

**Figure 8. Definition of RealTimeClass stereotype.**

# 5. From an UML-RTDB diagram to an Object Relationship schema

Many development tools are based on relational database [2]. Although the relational model is useful for many applications, we believe that it is not as well-suited as an object-oriented database model for applications that require complex data management, have complex relationships between data, first-class support for timing constraints, and more scheduling flexibility than serializability can provide. The mapping of an UML-RTDB class generates, in the relational model, a large number of tables. So, queries need many joins to retrieve needed data, and their execution becomes expensive (i.e. a large amount of time). The main reason of this inadequacy is that relational model can deal only with simple data. In other words, a relational model uses a tabular representation of the real-world entities in the *first normal form* (1NF). For instance, the mapping of the *Aircraft* class gives seven tables: one to represent the atomic attributes of the *Aircraft*, and six tables representing every object connected with the *Aircraft*: *Direction*, *Location*, *Altitude*, *Speed*, *Path* and *Lane*. Each table contains a *foreign key* referring to the *primary key* of the *Aircraft*. The selection of all information of an *Aircraft* needs six join operations. This decreases significantly the performance of the system.

For these reasons, we see that the mapping of real-time class to the object-relational model is more suitable than the relational model. In fact, the object-relational model allows the use of simple or complex structures. Each structure is defined through an appropriate mechanism, called **U**ser **D**efined **T**ypes (*UDT*). An object-relational table is thus defined by means of either complex or simple data. The recordings of an object-relational table represent concrete objects, which have methods endowed with an **O**bject **Id**entity (*OID*).

Moreover, object-relational technology is a relational technology which is extended with new capabilities, such as *methods*, *UDT*, etc. It offers two advantages: firstly, it is compatible with relational technology and provides a better support for complex object. Secondly, object-relational databases are becoming commonplace because many commercial **D**ata**B**ase **M**anagement **S**ystem (*DBMS*) are adding object-oriented capabilities to their products, such as *Oracle 11g*, *IBM DB2* and *PostgreSQL*. For These reasons, we base our work on the object-relational databases design.

To map real-time classes in the object-relational model, we proceed in the following manner.

## 5.1 Mapping of derived attributes
Derived attributes are mapped through the following actions:

*(1) Action 1:* Creation of an *UDT*, named *RealTime*, which contains three fields *CV*, *TS* and *VD* (cf. Query1).

*Query 1 Creation of an UDT for derived attributes*
```
SQL create type RealTime as object
  2. (Value number,
  3. TimeStampValue timestamp,
  4. ValidityDuration number)
  5. not final;
  6. /
```

*(2) Action 2:* For every attribute whose multiplicity is greater than one, we :
- Create an *UDT* which represents a nested table, named *NT_RealTime*, of *RealTime* type, when the exact value of the multiplicity was not mentioned.
- Create an *UDT* which represents an array, named *ARR_RealTime*, of *RealTime* type, when the exact value of the multiplicity was mentioned.

## 5.2 Mapping of sensor attributes
Sensor attributes are mapped through the following actions:

*(1) Action 1:* Creation of an UDT, named RealTimeSensor, composed of four fields: CV, TS, VD and MDE (cf. Query 2).

*Query 2 Creation of an UDT for sensor attributes*
```
SQL create type RealTimeSensor under RealTime
  2. (MaximumDataError number)
  3. /
```

*(2) Action 2:* Is the same as action 2 for the derived attribute mapping (cf. section 5.1).

## 6. Mapping of real-time class

Real-time class is mapped through the following actions:

*(1) Action 1:* Creation of an *UDT*, named *NOM_CTR_TYPE*, which contains the following fields (cf. Query 3):

*Query 3  Creation of an UDT for aircraft class*

```
SQL create type Aircraft as object
  2. (Identifier varchar2(15),
  3. Destination varchar2(15)
  4. Direction RealTimeSensor,
  5. Location RealTimeSensor,
  6. Path RealTime,
  7. Lane RealTime)
  8. /
```

- Classical attributes of the real-time class.
- Sensor attributes of the real-time class, with the suitable types (types obtained from the mapping of sensor attributes: AN_TYPE, NT_AN_TYPE or ARR_AN_TYPE).
- Derived attributes of the real-time class, with the appropriate types (types obtained from the mapping of the derived attributes: *AN_TYPE*, *NT_AN_TYPE* or *ARR_AN_TYPE*).

*(2) Action 2:* Creation of an object-relational table which has the same name as the real-time class. Then, we add other constraints (*primary key*, *foreign key*, etc.) (cf. Query 4).

*Query 4  Creation of an UDT for aircraft class*

```
SQL create table AircraftTable of
Aircraft
  2. (constraint    pk_AircraftTable
     primary key (Identifier));
```

Figure 9 illustrates the structure of an *Aircraft* real-time class. It encapsulates classical attributes, and real-time sensor and derived attributes.

## 7. Implementation

The implementation of our UML-RTDB components is done through an extension of an UML CASE Tool, named Fujaba (**F**rom **U**ML to **J**ava **A**nd **B**ack **A**gain). The Fujaba environment aims to provide round-trip engineering support for UML and Java. The main distinction to other UML tools is its tight integration of UML class and UML behaviour diagrams to a visual programming language. This integration enables Fujaba to perform a lot of static analysis work, facilitating the creation of a consistent overall specification. In addition, it turns these UML diagrams into a powerful visual programming language and allows covering the generation of complete application code. Since Fujaba is open source, we could add to it the appropriate tools and make it able to accept real-time database specification. Figure 10 shows the class diagram under Fujaba of our air traffic control application. We have chosen a *"Watch"* icon to indicate sensor attribute and a "Calculator" icon to indicate derived attribute of Aircraft real-time class. In addition, a *"SPO"* icon is used to indicate sporadic methods and *"PER"* icon to indicate periodic methods. Figure 12 presents Java code generated from the Aircraft real-time class. Figure 11 presents SQL queries generated also from the Aircraft real-time class.

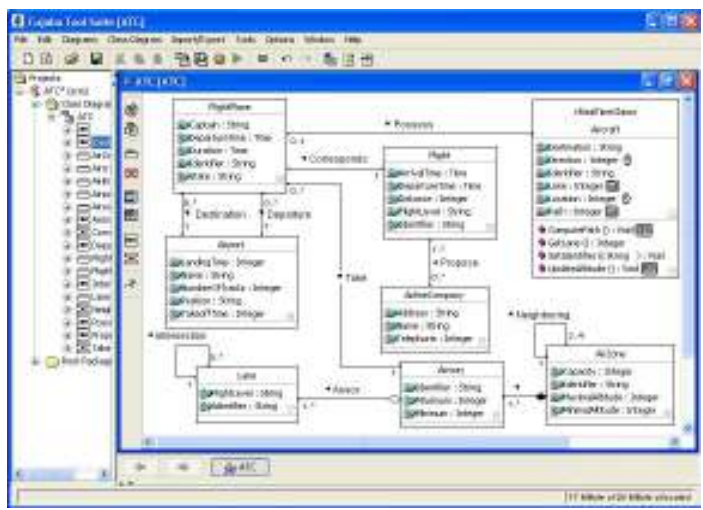

**Figure 9. Definition of RealTimeClass stereotype.**

**Figure 10. Air traffic control class diagram.**

**Figure 11. Java code generation from Aircraft real-time class.**

**Figure 12. SQL queries generation from Aircraft real-time class.**

## 8. Conclusion and future work

Many real-time applications need a database environment. But classical databases can not satisfy all requirements of these applications. We have studied, in this paper, the temporal requirements of real-time databases, for data and operations. To specify dynamic semantics and complex data structure of these databases, we have proposed a set of appropriate concepts and tools, giving a real-time object-oriented data model. This model incorporates new types of data (sensor attributes and derived attributes), time-constrained data, time-constrained methods (periodic methods, sporadic methods and aperiodic methods), and a concurrency control protocol (local controller). The framework we have designed and implemented to support this model helps designers to produce real-time applications, with temporal data and transactions semantics. It is composed by an UML profile for real-time databases, named UML-RTDB, a translator to a object-relational model, and an UML CASE Tool. UML-RTDB, based on UML2.2.1 Profiles package, contains a set of stereotypes expressing sensor attributes, derived attributes, periodic methods, sporadic methods and real-time class. So, it allows to design class diagrams for real-time databases. The translator is based on a set of mapping rules from a real-time class diagram to an object-relational model, which allows to use simple or complex structures. Finally, the UML CASE Tool is a support for the development of real-time databases. It is built as an extension of Fujaba, an open source standard UML CASE Tool.

In our future work, we will extend UML-RTDB with other stereotypes in order to express time-constrained associations and time-constrained multiplicities. Among them, we will study how to model dynamic aspects of real-time databases in the behavioural model (*Activity diagram* and *State diagram*). Moreover, we will add tools to Fujaba in order to support these dynamic and behavioural aspects. We note that in our laboratories (LITIS and MIRACL); there is a PhD which has just begun treating real-time object-oriented data model implementation, Concurrency Control Techniques and corresponding performance study.

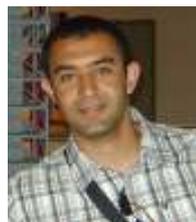

**Nizar Idoudi** is a PhD student at the Faculty of Sciences and Technics (University of Le Havre, France). His researches are done in two laboratories: MIRACL research group at ISIMS School (Tunisia) and LITIS laboratory (University of Le Havre). He received his Master degree in Computer Sciences from this university in 2004. His research interests include Modeling and design of Embedded Real-time systems and Real-time databases. He is particularly interested in UML Modeling Language and UML CASE Tools adapted to real-time context.

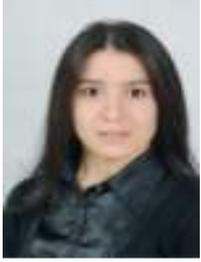

**Nada Louati** received the B.S and M.S degree from the University of Sfax, Tunisia, in 2006 and 2008 respectively. She is a Ph.D student at the department of Computer Science of the Faculty of Economics and Management of Sfax, Tunisia. Her Current research interests include real-time object modeling techniques for designing real-time database.

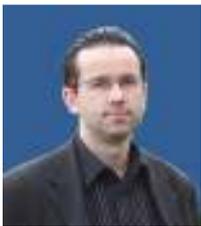

**Claude Duvallet** has obtained his PhD in October 2001. Since September 2003, he has been an associate professor at the University of Le Havre. His main topics of research are Real-Time Databases, Real-Time Systems and Multimedia Systems. He supervises many PhD students in these areas.

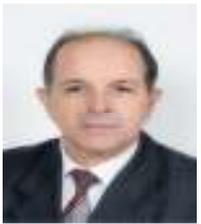

**Rafik Bouaziz** was a consulting Engineer in the Organization and Computer Science and a head of the department of Computer Science at CEGOS-TUNISIA between 1979 and 1986. Then, he integrated higher education area as an associate Professor in the Faculty of Economics and Management of Sfax (Tunisia). During the preparation of his PhD, he worked on temporal data management and historical record of data in Information Systems. Currently, his main research topics of interest are temporal databases, real-time databases, information systems, data warehousing and workflows.

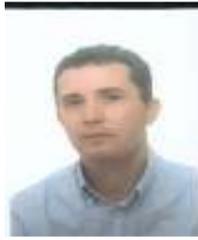

**Bruno Sadeg** is an associate Professor in the University of Le Havre (France). He is a head of a research team whose members work about *"Intelligent Transport Systems"*. He is particularly interested by real-time mechanisms in sensor databases, embedded in vehicles and/or databases located in some sites where vehicles get information, sometimes continuously.

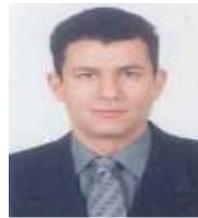

**Faïez GARGOURI** is a Professor at the Higher Institute of Computer Science and Multimedia School of Sfax (Tunisia), where he is the head. He obtained his master degree in computer science from the Paris-6 University in 1990, and a PhD from Paris-5 University in 1995. His research interest focus of the Information Systems field: Design, Quality Measurement, verification, data warehousing, multimedia, Knowledge Management, Ontology,… He published more than 50 papers in journals, international conferences and he is a PC member of multiple international conferences.